\documentclass[journal]{IEEEtran}
\usepackage{bm}
\usepackage{cite}
\usepackage{graphicx}
\usepackage{amsmath}
\usepackage{citesort}
\usepackage{amsfonts}
\usepackage{subfigure}
\usepackage{algorithmic}
\usepackage{algorithm}
\usepackage{amsmath,amssymb,verbatim,cite,citesort,amsopn,graphicx,citesort,url,color}
\usepackage{algorithm}
\usepackage{algorithmic}
\usepackage{multicol}
\usepackage{epsfig}
\usepackage{epstopdf}
\usepackage{balance}
\usepackage{stfloats}

\newtheorem{corollary}{\textbf{Corollary}}
\newtheorem{theorem}{\textbf{Theorem}}

\newtheorem{proposition}{\textbf{Proposition}}

\newtheorem{lemma}{\textbf{Lemma}}

\begin{document}

\begin{small}\title{{{Probabilistic Accumulate-then-Transmit in Wireless-Powered Covert Communications}}}\end{small}

\author{ { Yida Wang, Shihao Yan, \IEEEmembership{Member, IEEE,} Weiwei Yang, Caijun Zhong, \IEEEmembership{Senior Member, IEEE,}\\
and Derrick Wing Kwan Ng, \IEEEmembership{Fellow, IEEE}   }\\
\vspace{-0.15cm}
  \vspace{-1mm}





\vspace*{-8pt}
}

\maketitle

\begin{abstract}
In this paper, we investigate the optimal design of a wireless-powered covert communication (WP-CC) system, in which a probabilistic accumulate-then-transmit (ATT) protocol is proposed to maximize the communication covertness subject to a quality-of-service (QoS) requirement on communication. Specifically, in the considered WP-CC system, a full-duplex (FD) receiver transmits artificial noise (AN) to simultaneously charge an energy-constrained transmitter and to confuse a warden's detection on the transmitter's communication activity. With the probabilistic ATT protocol, the transmitter sends its information with a prior probability, i.e., $p$, conditioned on the available energy being sufficient. Our analysis shows that the probabilistic ATT protocol can achieve higher covertness than the traditional ATT protocol with $p=1$. In order to facilitate the optimal design of the WP-CC system, we also derive the warden's minimum detection error probability and characterize the effective covert rate from the transmitter to the receiver to quantify the communication covertness and quality, respectively. The derived analytical results facilitate the joint optimization of the probability $p$ and the information transmit power. We further present the optimal design of a cable-powered covert communication (CP-CC) system as a benchmark for comparison. Our simulation shows that the proposed probabilistic ATT protocol (with a varying $p$) can achieve the covertness upper bound determined by the CP-CC system, while the traditional ATT protocol (with $p=1$) cannot, which again confirms the benefits brought by the proposed probabilistic ATT in covert communications.
\end{abstract}

\begin{IEEEkeywords}
Covert communications, wireless-powered communications, artificial noise, transmit probability.
\end{IEEEkeywords}


\section{Introduction}\label{sec:1}

\subsection{Background}\label{sec:1-1}

With the roll-out of the Internet-of-things (IoT), there are numerous emerging wireless applications in everyday life \cite{IoT_Survey}. Unfortunately, the broadcast nature of radio propagation poses a severe threat to the security and privacy of wireless communications, which is one of the biggest obstacles to the widespread of IoT \cite{IoT_Security_Survey}. Against this background, conventional cryptographic and physical layer security technologies exploit encryption/decryption and the random nature of wireless medium, respectively, to offer the protection of communication content from eavesdropping ~\cite{PLS_multiple_antenna}. However, there exist many practical scenarios where the communication parties wish to further hide the communication activity from detection~\cite{Bash_Survery}. For example, a senior official with an embedded medical device desires covert information uploading in order to avoid the exposure of health privacy. In this case, an efficient design for covert communications is desired.

As a promising solution, covert communication aims to enable a communication between two nodes while guaranteeing a negligible detection probability of this communication activity at a warden \cite{Yan_Survery}. 
Inspired by this, a square root law was derived in \cite{Bash} by considering additive white Gaussian noise (AWGN) channels, which stated that no more than $\mathcal{O}(n)$ bits could be transmitted to a legitimate receiver reliably and covertly in $n$ channel uses. Since then, following \cite{Bash}, many researchers have examined various possible uncertainties at the warden's detection to achieve covertness. These include the uncertainties on the knowledge of the synchronization information \cite{Bash_Time}, the background noise \cite{Lee,He_Biao,new_channel_uncertainty}, and the additional interference \cite{Cognitive,Relay,zheng2019multi,Uninformed_Jammer,fullduplex,shu_feng,Backscatter,like2020optimal}. Specifically, \cite{Cognitive} exploited the channel uncertainty of an additional public link to shield the communication activity. Also, the authors in \cite{Relay} focused on a dual-hop relaying network, where the relay exploited the channel uncertainty of the second hop to transmit its own information to the receiver covertly. In addition, \cite{zheng2019multi} studied the interference uncertainty in the scenario with multiple interferers. {To further improve the communication covertness}, \cite{Uninformed_Jammer} introduced a dedicated jammer to generate artificial noise (AN) with {time} varying transmit power across multiple time slots. Furthermore,
\cite{fullduplex} conducted covert communications by exploiting AN generated from a full-duplex (FD) receiver, which can mitigate self-interference by applying advanced interference cancellation techniques. Then, the work on covert communications with a FD receiver was extended to different scenarios, such as a system with a finite number of channel uses \cite{shu_feng}, the backscatter communications \cite{Backscatter}, and a system adopting channel inverse power control~\cite{like2020optimal}. We note that the techniques developed in aforementioned works were based on one overly optimistic assumption, where a perpetual energy supply was available to the system for supporting continuous operations. Unfortunately, IoT devices are usually battery powered with a limited energy storage and their distributed deployments make conventional cable-type energy replenishment method impractical. In other words, the results of \cite{Cognitive,Relay,zheng2019multi,Uninformed_Jammer,fullduplex,shu_feng,Backscatter,like2020optimal} may not be applicable to some IoT applications in practical scenarios {and there is a need for IoT-oriented covert communications.}

In practice, IoT devices are often deployed in an environment that is hard to access, such as embedded medical sensors in a human body or reconnaissance sensors in a battlefield. Thus, it is challenging to enable {sustainability for IoT devices}. An emerging technology that can overcome this limitation is wireless power transfer (WPT), where wireless nodes can charge their batteries by harvesting energy from the received electromagnetic waves \cite{WPT_Survey,HTT,JiangXin,EH_Relay,Relay_new_accumulation}. Compared with the conventional cable-type energy supply, WPT is a promising paradigm to provide remote and controllable energy supply \cite{WPT_Survey}. In this context, \cite{HTT} investigated wireless-powered communications with a hybrid access point (HAP) and proposed the harvest-then-transmit (HTT) protocol, i.e., a user first harvests energy broadcast by a HAP and then transmits its information back to the HAP. In addition, \cite{JiangXin} introduced a separate friendly node to improve the physical layer security performance of the HTT protocol, where the friendly node played a dual-role of jammer and {power beacon}. Yet, the conventional HTT protocol follows a regular routine of operations for transmission, which means the communication activity is predictable and the communication covertness is hard to be guaranteed. Recently, \cite{EH_Relay} investigated covert communications in a simultaneous wireless information and power transfer (SWIPT) aided dual-hop relaying network, where the communication covertness is enabled by exploiting the channel uncertainty of the second hop. Furthermore, \cite{Relay_new_accumulation} considered the FD relaying protocol in SWIPT-aided covert communications. Despite the fruitful results in the literature, applying the approaches developed in the existing works to practical IoT scenarios still has several challenges as below. Firstly, the SWIPT technology is often adopted to support downlink transmission, while the downlink traffic in IoT networks is much lighter than that of the uplink traffic \cite{uplink_larger_downlink}. Secondly, varying the energy conversion efficiency \cite{EH_Relay} or the information transmit power \cite{Relay_new_accumulation} across time slots is costly and hard to achieve for IoT devices with low-cost hardware. Thirdly, the possibility of energy accumulation multiple across time slots has been ignored, which is a vital energy replenishment for energy-constrained IoT devices. As such, a new framework for wireless-powered covert communications (WP-CC) is desired.
\subsection{Our Approach and Contribution}\label{sec:1-2}
In this work, we tackle the optimal design of a WP-CC system, where a FD receiver continuously transmits AN to simultaneously charge an energy-constrained transmitter and to confuse a warden's detection on this transmitter's communication activity. In the considered WP-CC system, the transmitter either accumulates the harvested energy or transmits information in a random manner with an optimized probability. We also examine the performance of a cable-powered covert communication (CP-CC) system as a benchmark, where the AN generated by the FD receiver only aims to shield the communication activity of the transmitter. The main contributions of this work are summarized as follows.

\begin{itemize}
 \item For the first time, we tackle the optimal design of the WP-CC system by proposing a probabilistic accumulate-then-transmit (ATT) protocol. In this probabilistic ATT protocol, the transmitter sends information with a probability of $p$ even when it has accumulated sufficient amount of energy for transmission, i.e., adopting the intermittent transmission even when the available energy at transmitter is sufficient. Then, considering the impact of WPT, we derive an explicit expression for the transmitter's prior transmit probability $q$ as a non-linear function of the information transmit power $P_s$ and the aforementioned conditional prior transmit probability $p$, in order to facilitate the optimal design of the WP-CC system.

\item In the considered WP-CC system, we analytically derive the warden's minimum detection error probability and the achievable effective covert rate from the transmitter to the receiver to evaluate the communication covertness and quality, respectively. Then, we formulate an optimization problem to determine the optimal information transmit power $P_s$ and the conditional prior transmit probability $p$ to maximize the communication covertness subject to a constraint on the communication quality. Our analysis reveals that varying $p$ can achieve higher communication covertness relative to setting $p=1$, which shows the benefits of using the proposed probabilistic ATT protocol in the WP-CC system.

\item We study the optimal design of the CP-CC system as a benchmark, in which the prior transmit probability $q=p$ is directly optimized together with the information transmit power $P_s$. Our analysis reveals that the achievable covertness in the CP-CC system serves as an upper bound on the one achieved in the WP-CC system. Meanwhile, we find that the WP-CC system with $p=1$ cannot generally achieve the covertness upper bound determined by the CP-CC system, while varying $p$ can approach the performance of the CP-CC system. This again demonstrates the superiority of the proposed probabilistic ATT protocol in the context of WP-CC (i.e., wireless-powered covert communications).
\end{itemize}

\subsection{Organization and Notation}\label{sec:1-4}
The rest of this paper is organized as follows. Section \uppercase\expandafter{\romannumeral2} details our system model and the adopted assumptions. Sections \uppercase\expandafter{\romannumeral3} and \uppercase\expandafter{\romannumeral4} present the analysis of the WP-CC and CP-CC systems, respectively. Section \uppercase\expandafter{\romannumeral5} provides numerical results to confirm the accuracy of our analysis. Section \uppercase\expandafter{\romannumeral6} draws conclusions.

\textit{Notation}: Scalar variables are denoted by italic symbols. Vectors are denoted by boldface symbols. Given a complex number, $\left|  \cdot  \right|$ denotes its modulus. Given an event, $\Pr \left(  \cdot  \right)$ denotes its probability of occurrence. $\mathbb{E}\left(  \cdot  \right)$ denotes the expectation operation. $\mathbb{C}^{a \times b}$ denotes the space of $a \times b$ complex-valued matrices.

\section{System Model}\label{sec:2}

\subsection{Considered Scenario and Adopted Assumptions}\label{sec:2-1}

\begin{figure}
\begin{center}
  \includegraphics[width=3.5in]{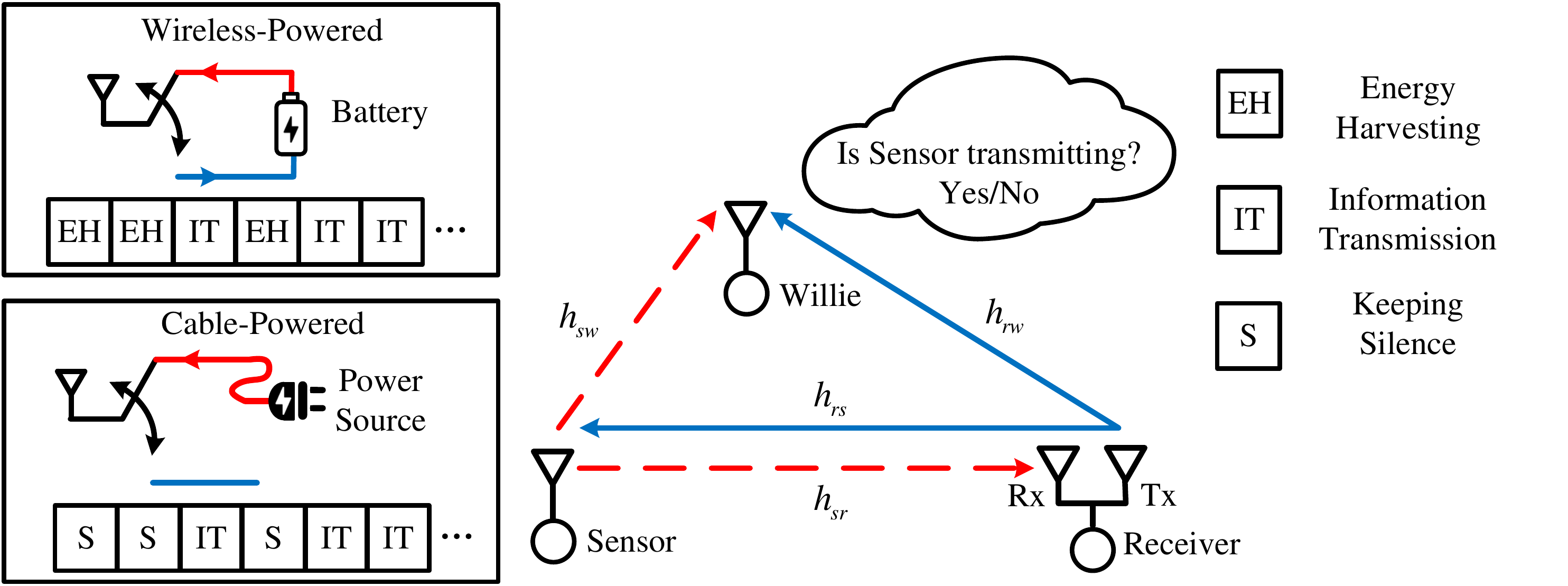}\\
  \caption{The system model of covert communications with a FD receiver. When Sensor does not transmit information, Sensor harvests energy from AN generated by the FD receiver in the WP-CC system, while Sensor keeps silence in the CP-CC system. }\label{system model}
\end{center}
\end{figure}

We consider a covert wireless communication system as shown in Fig. 1, where a transmitter (Sensor) tries to transmit its information to a FD receiver (Receiver) with a prior transmit probability under the supervision of a warden (Willie). Both Sensor and Willie are single-antenna devices. Meanwhile, besides a receiving antenna, Receiver uses an additional antenna to transmit AN to confuse Willie, who tries to detect whether there is a transmission from Sensor to Receiver \cite{fullduplex,shu_feng,Backscatter,like2020optimal}. We assume that the duration of each time slot is $T$ and each time slot contains $n$ channel uses. The channels of Sensor-Willie, Sensor-Receiver, Receiver-Sensor, and Receiver-Willie are denoted as $\left\{ {{h_{sw}}{\rm{,}}{h_{sr}}{\rm{,}}{h_{rs}}{\rm{,}}{h_{rw}}} \right\} \in \mathbb{C}$, respectively. In addition, we assume that the distances of Sensor-Receiver and Receiver-Willie channels are relatively large \cite{Cognitive,Relay,zheng2019multi,Uninformed_Jammer,fullduplex}. Thus, $h_{sr}$, $h_{rs}$, and $h_{rw}$ are modeled as quasi-static Rayleigh fading channels \cite{Wireless_Book} and the mean value of ${\left| {{h_{xy}}} \right|^2}$ over different slots is denoted by $1/{\lambda _{xy}}$, where $xy \in \left\{ {sw{\rm{,}}\;sr{\rm{,}}\;rs{\rm{,}}\;rw} \right\}$. In addition, the channel gain of $h_{sw}$ is denoted as ${\left| {{h_{sw}}} \right|^2}$ known by Willie, which will be further explained in Section \ref{sec:2-3} by considering a worst-case scenario from a conservative point of view.

If Sensor has sufficient energy and decides to transmit information, the signal received at Receiver in each channel use is given by
\begin{equation}\label{eqq12}
y_r = \sqrt {{P_s}}{h_{sr}}{x}_s + \sqrt {\phi  {P_r}}{{x}}_r+{n}_r,
\end{equation}
where $P_s$ is the information transmit power, $x_s \in \mathbb{C}$ is the information signal transmitted by Sensor satisfying $\mathbb{E}\left( {{{\left| {{x_s}} \right|}^2}} \right){\rm{ = }}1$, $P_r$ is the AN transmit power,  $x_r \in \mathbb{C}$ is the AN signal transmitted by Receiver satisfying $\mathbb{E}\left( {{{\left| {{x_r}} \right|}^2}} \right){\rm{ = }}1$, and $n_r\in \mathbb{C}$ is the AWGN at Receiver with zero mean and variance $\sigma _r^2$. Since the AN signal is known to Receiver, the self-interference caused by AN can be suppressed by interference cancellation techniques \cite{Huguojie1,Huguojie2}. In this work, we assume that the self-interference cannot be perfectly cancelled and the residual self-interference is $\phi P_r$, where $\phi$ is the self-interference cancellation coefficient capturing the quality of cancellation. It is noted that $\phi=0$ refers to the ideal case with perfect self-interference cancellation, while $0 < \phi  \le 1$ corresponds to different self-interference cancellation capabilities \cite{imperfect_SIC}. As per \eqref{eqq12}, the signal-to-interference-plus-noise ratio (SINR) at Receiver in each channel use is given by
\begin{align}
{\gamma _r}=\frac{{{P_s}{{\left| {{h_{sr}}} \right|}^2}}}{\phi P_r+{\sigma _r^2}}.
\end{align}

In this work, we consider a fixed-rate transmission from Sensor to Receiver, where $R$ is the predetermined transmission rate. Due to the random nature of $h_{sr}$, a communication outage from Sensor to Receiver occurs when $C \le R$, where $C={\log _2}\left( {1+{\gamma _r}} \right)$ is the channel achievable rate from Sensor to Receiver. Besides the transmission outage probability, the prior transmit probability $q$ also affects the amount of information that can be reliably transmitted from Sensor to Receiver. As such, in this work, we adopt the effective covert rate to evaluate the covert communication quality from Sensor to Receiver \cite{fullduplex}, which is defined as
\begin{align}\label{addd11}
{R_c} = qR\left( {1 - {P_\mathrm{out}}} \right),
 \end{align}
where $P_\mathrm{out}=\Pr(C<R)$ is the transmission outage probability.

\subsection{Available Energy at Sensor} \label{sec:2-2}

\begin{figure}
\begin{center}
  \includegraphics[width=3.5in]{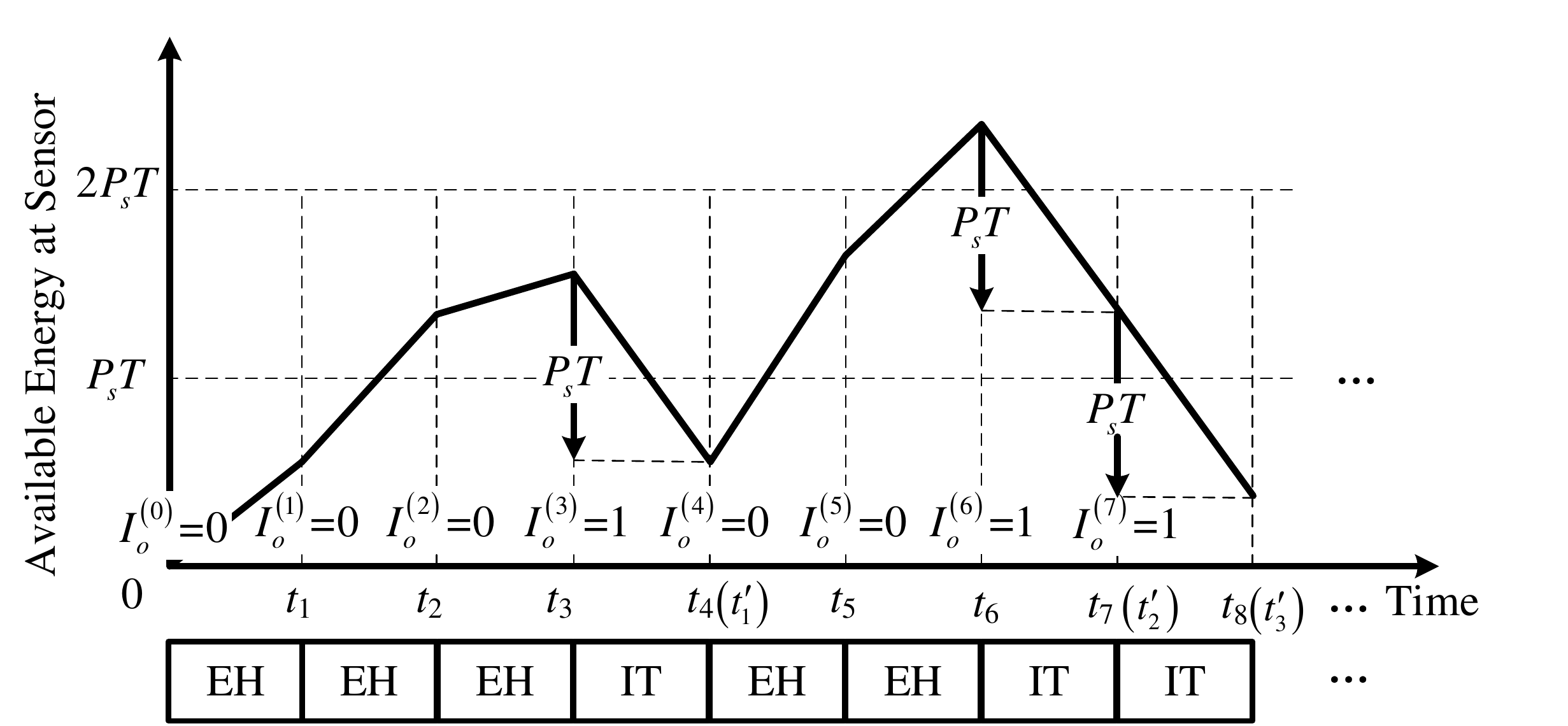}\\
  \caption{An example of energy evolution at Sensor in the WP-CC system.}\label{system model}
\end{center}
\end{figure}

\subsubsection{Wireless-Powered Communications}\label{sec:2-2-1}
Sensor is generally energy-constrained. Thus, WPT is a promising solution to achieve controllable power supply. Specifically, we adopt the ATT protocol (i.e., the accumulate-then-transmit protocol), where each time slot is devoted entirely
either to energy harvesting (EH) or information transmission (IT) \cite{Bi_Ying}. Compared with the HTT protocol \cite{HTT}, i.e., switching the operation between EH and IT within every time slot, the ATT protocol enables continuous EH for accumulating energy over long period of time, which relaxes the constraint on the available energy that devices
can use. In practice, within an EH time slot, Sensor does not transmit information, but receives signals from the FD Receiver for EH. The signal received at Sensor in each channel use is given by
\begin{equation}\label{add99}
{{{y}}}_s{\rm{ = }}\sqrt {{P_r}} {h_{rs}}{{x}}_r + {{{n}}_s},
\end{equation}
where $n_s\in \mathbb{C}$ is the AWGN at Sensor with zero mean and variance $\sigma _s^2$. In contrast, within an IT time slot, Sensor does transmit information and the required energy is $P_sT$. Then, we denote the start time of the $k$-th time slot as $t_k=kT$, where $k \in {\rm{\{ 0,1,2,3}} \cdots {\rm{\} }}$. As such, the available energy at $t_k$ is denoted as $E_a^{\left( k \right)}$. However, it is noted that $E_a^{\left( k \right)}$ is not always necessarily sufficient to support IT. Thus, we need to further analyze the evolution of $E_a^{\left( k \right)}$ across time {as will be discussed in this section later.}

In the proposed design, Sensor cannot transmit information when $E_a^{\left( k \right)}<P_sT$. Meanwhile, we consider that Sensor does not always transmit information even when $E_a^{\left( k \right)}\ge P_sT$ holds, but transmits information with a conditional prior probability of $p>0$. It means that Sensor deliberately harvests energy with probability $1-p$ even when $E_a^{\left( k \right)}\ge P_sT$ holds (i.e., the probabilistic ATT protocol).

Then, we introduce an indicator function to illustrate the operation of Sensor at the following time slot of $t_k$, which is given by
\begin{equation}
I_o^{\left( k \right)} \!=\! \left\{ \begin{array}{l}
0,\;\;\;\;\;\;\;\;\;\;\;\;\;\;\;\;\;\;\;\;\;\;\;
\;\;\;\;\;\;\;\;\;\;\;\;\;\;\;\;\;E_a^{\left( k \right)} < P_sT ,\\
\left. \!\!\!\begin{array}{l}
0,\;\text{with\;probability} \;{1 - p}\;\;\\
1,\;\text{with\;probability} \;p
\end{array} \!\!\!\right\},E_a^{\left( k \right)} \ge P_sT ,
\end{array} \right.
\end{equation}
where $I_o^{\left( k \right)}=0$ means that this time slot is an EH time slot and $I_o^{\left( k \right)}=1$ means that this time slot is an IT time slot. In order to guarantee IT from Sensor, the embedded battery in Sensor is set sufficiently large to accumulate a significant amount of energy \cite{Nasir}. As such, the energy evolution of Sensor is given by
\begin{equation}
E_a^{\left( {k + 1} \right)} = E_a^{\left( k \right)} + \left( {1 - I_o^{\left( k \right)}} \right)E_h^{\left( k \right)} - I_o^{\left( k \right)}{P_s}T, \forall k,
\end{equation}
where $E_h^{\left( k \right)}$ denotes the harvested energy if the following time slot of $t_k$ is an EH time. Based on the assumption that $P_r$ is sufficiently large such that the energy harvested from the AWGN is negligible \cite{HTT}, $E_h^{\left( k \right)}$ is given by
\begin{equation}\label{eqq31}
E_h^{\left( k \right)}{\rm{ = }}\eta {P_r}{\left| {{h_{rs}}} \right|^2}T,
\end{equation}
where $0<\eta<1$ is the energy conversion efficiency at Sensor. Following the existing works \cite{WPT_Survey,HTT,JiangXin,EH_Relay,Relay_new_accumulation}, we consider a linear EH model with perfect CSI $h_{rs}$ at Sensor for analytical tractability. It is noted that the linear EH model is accurate when the input power is below a certain threshold, which is suitable for our application. Furthermore, a more general scenario with a non-linear EH model \cite{non_linear} and imperfect CSI \cite{imperfect_EH} is worth studying in future works.

We present one example of energy evolution within $8$ consecutive time slots in Fig. 2 for illustration of the probabilistic ATT protocol. Here, we define some auxiliary variables to simplify the following analysis. Specifically, the end time of the $l$-th IT operation is denoted as $t'_l$,  where $l \in {\rm{\{ 1,2,3}} \cdots {\rm{\} }}$. The total number of past time slots before $t'_l$ is denoted as $K_l$, which is given by
\begin{equation}
{K_l} = \max \left( {k\;|\;  {t_k} \le \;t'_l\; } \right).
\end{equation}
Thus, the number of successive EH time slots before $t'_{l}$ is given by
\begin{equation}
{M_l}{\rm{ = }}\left\{ \begin{array}{l}\label{add1}
{K_1} - 1{\rm{,}}\;\;\;\;\;\;\;\;\;\;\;\;\;\;\;l{\rm{ = }}1,\\
{K_l} - {K_{l - 1}} - 1{\rm{,}}\;\;\;l{\rm{ > }}1,
\end{array} \right.
\end{equation}
where ${M_l} \in {\rm{\{ 0,1,2,3}} \cdots {\rm{\} }}$. Furthermore, we define the number of successive EH time slots before $t'_{l}$ with $E_a^{\left( M_l \right)} < P_sT $ and with $E_a^{\left( M_l \right)} \ge P_sT $ as $M_l'$ and $N_l$, respectively. As such, $M_l$ as a function of $M_l'$ and $N_l$ is given by
\begin{equation}
{M_l}{\rm{ = }}\left\{ \begin{array}{l} \label{add2}
{N_l},\;\;\;\;\;\;\;\;\;\;\;\;\;\;\;\;\;\;\;\;\;{M_l'}= 0,\\
{M_l'}{\rm{ + }}1{\rm{ + }}{N_l},\;\;\;\;\;\;\;\;\; {M_l'} \ge {\rm{1}}.
\end{array} \right.
\end{equation}
It is noted that $M_l=0$ means that there is no EH time slot before $t'_l$ and $M_{l+1}'=0$ means that there is sufficient energy after $t_{l}'$, such as $M_3=M_3'=0$ in Fig. 2.

\subsubsection{Cable-Powered Communications}\label{sec:2-2-2}
In contrast to wireless-powered communications, most works on covert communications considered that the available energy of all nodes was sufficient at any time slot \cite{Bash,Lee,He_Biao,Bash_Time,new_channel_uncertainty,Cognitive,Relay,zheng2019multi,Uninformed_Jammer,fullduplex,shu_feng,Backscatter,like2020optimal}. It can be easily achieved by plugging a cable from a power source to a communication node as shown in Fig.~1. For such cable-powered communications,
there is always sufficient energy for transmission and thus we have
\begin{equation}\label{add29}
q = p,
\end{equation}
where $q$ is the prior transmit probability for Sensor to conduct covert communications and $p$ is the conditional prior transmit probability, which is the prior probability of covert transmission conditioned on that Sensor has sufficient energy for IT.

It is noted that Sensor in cable-powered communications switches its operation between IT and keeping silent across time slots. As such, there exists an affine mapping from $p$ to $q$, while such a mapping is non-linear in the WP-CC system due to possible energy insufficiency.

\subsection{Activity Detection at Willie} \label{sec:2-3}

In the considered system, Sensor and Receiver work cooperatively to fight against Willie. As such, we assume that Receiver can acquire the estimated channel state information (CSI) based on secure feedback links, whereas it is impossible for Willie to obtain the estimated CSI without such feedback links \cite{EH_Relay,D2D}. Therefore, Willie has its uncertainty on the instantaneous CSI of $h_{rw}$. From a conservative point of view, we still consider that Willie knows the instantaneous CSI of $h_{sw}$, the statistical CSI of $h_{rw}$, the conditional prior transmit probability $p$ in the WP-CC system, and the prior transmit probability $q$ in the CP-CC system. Due to the low-cost hardware in IoT devices, varying the transmit power across time slots is costly and hard to achieve \cite{shu_feng}. Thus, we assume that the transmit power at Sensor and Receiver are fixed and public, i.e., $P_s$ and $P_r$ are known to Willie.

In this context, Willie needs to perform a hypothesis testing based on its observation within a time slot \cite{Bash}, in which Sensor does not transmit in the null hypothesis $H_0$, but transmits in the alternative hypothesis $H_1$. Specifically, the composite received signal at Willie in each channel use is given by
\begin{equation}\label{received_willie}
{{{y}}_w}=\left\{ \begin{array}{l}
\sqrt {{P_r}} {h_{rw}}{{x}}_r + {{{n}}_w},\;\;\;\;\;\;\;\;\;\;\;\;\;\;\;\;\;\;\;\;\;\;\;\;{H_0},\\
\sqrt {{P_r}} {h_{rw}}{{x}}_r \!+\! \sqrt {{P_s}} {h_{sw}}{{x}}_s \!+\! {{{n}}_w}{\rm{,}}\;\;\;\;\;\; {H_1},
\end{array} \right.
\end{equation}
where ${n}_w \in \mathbb{C}$ is the AWGN at Willie with zero mean and variance $\sigma _w^2$.

The detection performance of Willie is normally measured by
the detection error probability \cite{Test}, which is defined as
\begin{equation}\label{eqq6}
\xi  = (1-q){P_\mathrm{FA}} + q{P_\mathrm{MD}},
\end{equation}
where $1-q = {\rm{Pr}}\left( {{H_0}} \right)$ and $q = {\rm{Pr}}\left( {{H_1}} \right)$ are the prior probabilities of hypotheses $H_0$ and $H_1$, respectively, ${P_\mathrm{FA}} = \Pr \left( {{D_1}{\rm{|}}{H_0}} \right)$ is the false alarm probability, ${P_\mathrm{MD}} = \Pr \left( {{D_0}{\rm{|}}{H_1}} \right)$ is the miss detection probability, and variables $D_1$ and $D_0$ are defined as the events that Willie makes decisions in the favor of Sensor transmitting or not, respectively.

\section{Wireless-Powered Covert Communications} \label{sec:3}

In this section, we tackle the optimal design of the WP-CC system. To this end, we first derive an expression for the prior transmit probability $q$ as a non-linear function of $P_s$ and $p$. Then, we analyze Willie's detection performance to evaluate the communication covertness. At last, we optimize the values of $P_s$ and $p$ to maximize the communication covertness under a constraint on the effective covert rate.

\subsection{The Prior Transmit Probability $q$} \label{sec:3-1}

Now, based on the energy evolution discussed in Section~\ref{sec:2-2}, we have to determine the distribution of $E_a^{\left( {{k}} \right)}$ and the distribution of $M'_l$ conditioned on $E_a^{\left( {{M_l}} \right)} \le P_sT$ in order to explicitly characterize $q$. To this end, we first present the following two lemmas to characterize the EH dynamic of the system.

\begin{lemma}
The probability density function (pdf) of the available energy at Sensor at the start of $t_k$, i.e., $E_a^{(k)}$, is given by
\begin{equation}\label{eqq3}
{f_{E_a^{(k)}}}\left( z \right){\rm{ = }}{\lambda _E}\exp \left( { - {\lambda _E}z} \right),\;{\kern 1pt} z > 0,
\end{equation}
where ${\lambda _E} = \frac{{{\lambda _{rs}}}}{{\eta {P_r}T}}$.
\end{lemma}

\begin{IEEEproof}
The detailed proof is presented in Appendix A.
\end{IEEEproof}

\begin{lemma}
When $E_a^{\left( {{M_l}} \right)} \le {P_sT}$, $M'_l$ is a Poisson random variable with mean $1/\lambda _{M'_l}$ and its probability mass function (pmf) is given by
\begin{equation}\label{eqq4}
\!\!p_{M'_l|E_a^{\left( {{M_l}} \right)}}  \left( {m| E_a^{\left( {{M_l}} \right)}  } \right) \!=\! \frac{{\lambda _{M'_l}^m\exp ( \!-\! {\lambda _{M'_l}})}}{{m!}},E_a^{\left( {{M_l}} \right)} \!<\! {P_sT},
\end{equation}
where ${\lambda _{{M'_l}}} = \frac{{{\lambda _{rs}}\left( {{P_s}T - E_a^{\left( {{M_l}} \right)}} \right)}}{{\eta {P_r}T}}$.
\end{lemma}

\begin{IEEEproof}
It is noted that $E_h^{(k)}$ follows the exponential distribution with mean $1/{\lambda _{{E}}}$, which has been proved in Appendix A. Therefore, according to \cite[Theorem 2.2.4]{Discrete_Book}, ${M'_l}$ follows a Poisson distribution with mean $\frac{{{P_s}T - E_a^{\left( {{M_l}} \right)}}}{{{\lambda _E}}}$, which leads to \eqref{eqq4}.
\end{IEEEproof}

Following Lemma 1 and Lemma 2, we can derive the exact expression for $q$ in the following theorem.

\begin{theorem}\label{theorem1}
 The prior transmit probability $q$ in the WP-CC system as a function of $P_s$ and $p$ is derived as
\begin{align}\label{add3}
q = \frac{1}{{1 + \frac{{{\rm{1}} - p}}{{{p^2}}}{\rm{ + }}\frac{{{\lambda _{rs}}P_s }}{\eta {P_r}}}}.
\end{align}
\end{theorem}

\begin{IEEEproof}
According to the definitions of $H_0$ and $H_1$, the prior transmit probability is given by
\begin{align}
q &=\Pr \left( {{H_1}} \right) =1-\Pr \left( {{H_0}} \right)=\frac{1}{{{1+\mathbb{E}_{{M_l}}}\left( {{M_l} } \right) }}. \label{eqq28}
\end{align}
Due to the intractability of ${\mathbb{E}_{{M_l}}}\left( {{M_l} } \right)$ defined in \eqref{add1}, as an alternative, we derive ${\mathbb{E}_{{M_l}}}\left( {{M_l} } \right)$ based on \eqref{add2}, which is given by
\begin{align}\label{eqq30}
{\mathbb{E}_{{M_l}}}\left( {{M_l}} \right) &= {\mathbb{E}_{E_a^{\left( {{M_l}} \right)}}}\left[ {{\mathbb{E}_{{M_l}{\rm{|}}E_a^{\left( {{M_l}} \right)}}}\left(  {{M_l}{\rm{|}}E_a^{\left( {{M_l}} \right)}} \right)} \right] \nonumber \\
& =\int_0^{{P_s}T} {{\mathbb{E}_{{M_l}{\rm{|}}E_a^{\left( {{M_l}} \right)}}}\left( {{M_l}^\prime  + 1{\rm{ + }}{N_l}} \right){f_{E_a^{\left( {{M_l}} \right)}}}\left( z \right)dz}  \nonumber\\
& + \int_{{P_s}T}^\infty  {{_{{M_l}{\rm{|}}E_a^{\left( {{M_l}} \right)}}}\left( {{N_l}} \right){f_{E_a^{\left( {{M_l}} \right)}}}\left( z \right)dz} \nonumber \\
& =\int_0^{{P_s}T} \mathbb{E}{\left[ {\frac{{{\lambda _{rs}}}}{{\eta {P_r}T}}\left( {{P_s}T - z} \right){\rm{ + }}1} \right]{f_{E_a^{\left( {{M_l}} \right)}}}\left( z \right)dz}   \nonumber \\
& + \sum\limits_{i{\rm{ = }}0}^\infty  {{{\left( {1 - p} \right)}^i}i} \nonumber \\
& \mathop {\rm{ = }}\limits^{(a)} \frac{{{\lambda _{rs}}{P_s}}}{{\eta {P_r}}} + \frac{{1 - p}}{{{p^2}}},
\end{align}
where (a) is based on \cite[Eq. 3.351.1]{Integral} and \cite[Eq. 0.231.2]{Integral}. As per \eqref{eqq28} and \eqref{eqq30}, we can acquire the expression for $q$ given in \eqref{add3}.
\end{IEEEproof}

Based on Theorem~\ref{theorem1}, we first note that $q$ is a monotonically decreasing function of $P_s$, which means that $q$ can be increased by reducing $P_s$. In addition, we note that the monotonicity of $q$ with respect to (w.r.t.) $p$ depends on the term of $\frac{{{\rm{1}} - p}}{{{p^2}}}$, of which the first derivative w.r.t. $p$ is given by
 \begin{align} \label{add98}
\frac{{\partial \frac{{1 - p}}{{{p^2}}}}}{{\partial p}} = \frac{{p - 2}}{{{p^3}}} < 0,\;\text{for}\;0 < p \le 1.
 \end{align}
As per \eqref{add98}, $q$ is a monotonically increasing function of $p$. Therefore, Sensor can adjust $P_s$ or $p$ to control $q$ in the WP-CC system. Furthermore, we note that the maximum value of $q$ is $\frac{1}{{1 + \frac{{{\lambda _{rs}}{P_s}}}{{\eta {P_r}}}}}$ achieved with $p=1$, which is still lower than $1$ due to the possible energy limitation. In particular, $p=1$ means that Sensor always transmits information if $E_a^{\left( k \right)} \ge {P_s}T$ holds, which maximizes the transmission throughput and thus have been widely used in the literature of WPT without considering covertness constraint (e.g. \cite{Krikidis_single_Relay,accumulate,Bi_Ying}). Note that as we will show in Section \ref{sec:3-3}, $p=1$ is not always the best strategy in the considered WP-CC system for guaranteeing communication covertness.




\subsection{Willie's Minimum Detection Error Probability} \label{sec:3-2}

In order to evaluate the communication covertness, we next analyze Willie's detection performance. As per the Neyman-Pearson criterion \cite[Theorem 3.2.1]{Test}, the optimal approach for Willie to minimize its detection error probability is to adopt the likelihood ratio test (LRT). {With the aid of the Neyman-Fisher factorization theorem \cite[Corollary 7.7.1]{NeymanFisher} and the likelihood ratio order concept \cite[Theorem 1.C.11]{StochasticOrders}}, the LRT in the considered system can be transformed into the test of the average power received within a time slot, which is given by
\begin{equation}\label{eqq7}
{P_w}\mathop {{\rm{\gtrless}}}\limits_{{D_{\rm{0}}}}^{{D_{\rm{1}}}} \tau,
\end{equation}
where ${P_w}$ is the average power received at Willie within a time slot, and $\tau$ is Willie's detection threshold.
Considering the case that the number of channel uses is sufficiently large (i.e., $n \to \infty $) and following \eqref{received_willie}, $P_w$ is given by
\begin{equation}\label{eqq8}
{P_w} = \left\{ \begin{array}{l}
{P_r}{\left| {{h_{rw}}} \right|^2} + \sigma _w^2,\;\;\;\;\;\;\;\;\;\;\;\;\;\;\;\;\;\;{H_0},\\
{P_r}{\left| {{h_{rw}}} \right|^2} + {P_s}{{\left| {{h_{sw}}} \right|^2}} + \sigma _w^2,\;{H_1}.
\end{array} \right.
\end{equation}

Following \eqref{eqq6} and \eqref{eqq7}, we present Willie's detection error probability in the following lemma.

\begin{lemma}\label{lemma3}
Willie's detection error probability is given by
\begin{align}\label{eqq10}
\xi  &= \left( {{\rm{1}} - q} \right){P_{{\rm{FA}}}} + q{P_{{\rm{MD}}}} \nonumber \\
&= \left\{ {\begin{array}{*{20}{l}}
{1 - q,\;\;\;{\kern 1pt} \;{\kern 1pt} \;{\kern 1pt} \;{\kern 1pt} \;{\kern 1pt} \;{\kern 1pt} \;{\kern 1pt} \;{\kern 1pt} \;{\kern 1pt} \;{\kern 1pt} \;{\kern 1pt} \;{\kern 1pt} \;{\kern 1pt} \;{\kern 1pt} \;{\kern 1pt} \;{\kern 1pt} \;{\kern 1pt} \;{\kern 1pt} \;{\kern 1pt} \tau  \le \sigma _w^2,}\\
{\left( {1 - q} \right){\varphi _1},\;\;\;\;{\kern 1pt} \;{\kern 1pt} \;{\kern 1pt} \;{\kern 1pt} \;{\kern 1pt} \;{\kern 1pt}  \sigma _w^2 < \tau  \le \sigma _w^2 + {P_s}{{\left| {{h_{sw}}} \right|}^2},}\\
{\left( {1 - q} \right){\varphi _1} - q\left( {1 - {\varphi _2}} \right),\tau  > \sigma _w^2 + {P_s}{{\left| {{h_{sw}}} \right|}^2},}
\end{array}} \right.
\end{align}
where
\begin{align}
{\varphi _1} &= \exp \left( { - \frac{{{\lambda _{rw}}\left( {\tau  \!-\! \sigma _w^2} \right)}}{{{P_r}}}} \right), \nonumber \\
{\varphi _2} &= \exp \left( { - \frac{{{\lambda _{rw}}\left( {\tau  \!-\! \sigma _w^2 \!-\! {P_s}{{\left| {{h_{sw}}} \right|^2}}} \right)}}{{{P_r}}}} \right). \nonumber
\end{align}
\end{lemma}

\begin{IEEEproof}
With the aid of \eqref{eqq7} and \eqref{eqq8}, ${P_\mathrm{FA}}$ and ${P_\mathrm{MD}}$ can be derived as
\begin{align}
\!\!\!\!\!{P_\mathrm{FA}} &= \Pr \left( {{P_r}{{\left| {{h_{rw}}} \right|}^2} + \sigma _w^2 > \tau } \right) \nonumber \\
&= \left\{ \begin{array}{l}
1,\;\;\;\;\;\;\;\;\;\;\;\; \tau  \le \sigma _w^2,\\
 \varphi _1, \;\;\;\;\;\;\;\;\;\; \tau  > \sigma _w^2,
\end{array} \right.\\
 {P_\mathrm{MD}} &= \Pr \left( {{P_r}{{\left| {{h_{rw}}} \right|}^2} + {P_s}{{\left| {{h_{sw}}} \right|^2}}{\rm{ + }}\sigma _w^2{\rm{ < }}\tau } \right) \nonumber \\
&= \left\{ \begin{array}{l}
0, \;\;\;\;\;\;\;\;\;\;\;\; \tau \le \sigma _w^2 + {P_s}{{\left| {{h_{sw}}} \right|^2}},\\
1 - \varphi _2,\;\;\;\;\tau > \sigma _w^2 + {P_s}{{\left| {{h_{sw}}} \right|^2}}.
\end{array} \right.
\end{align}
Then, Willie's detection error probability is acquired as per \eqref{eqq6}.
\end{IEEEproof}

Lemma~\ref{lemma3} clearly shows that Willie's detection error probability highly depends on its adopted detection threshold $\tau$. From a conservative point of view, in covert communications we consider that Willie can acquire the prior transmit probability at Sensor and adopt the optimal detection threshold in the decision rule \eqref{eqq7} to achieve the minimum detection error probability, as detailed in the following theorem.

\begin{theorem}\label{theorem2}
Willie's optimal detection threshold in the detection rule \eqref{eqq7} is given by
\begin{equation}\label{eqq11}
{\tau ^\ast} = \left\{ {\begin{array}{*{20}{l}}
{ + \infty ,\;{\kern 1pt} \;{\kern 1pt} \;{\kern 1pt} \;{\kern 1pt} \;{\kern 1pt} \;{\kern 1pt} \;{\kern 1pt} \;{\kern 1pt} \;{\kern 1pt} \;{\kern 1pt} \;{\kern 1pt} \;{\kern 1pt} \;{\kern 1pt} q \le \frac{1}{{1 + \exp \left( {\frac{{{\lambda _{rw}}{{\left| {{h_{sw}}} \right|^2}}{P_s}}}{{{P_r}}}} \right)}},}\\
{\sigma _w^2 + {P_s}{{\left| {{h_{sw}}} \right|^2}},\;\; q > \frac{1}{{1 + \exp \left( {\frac{{{\lambda _{rw}}{{\left| {{h_{sw}}} \right|^2}}{P_s}}}{{{P_r}}}} \right)}},}
\end{array}} \right.
\end{equation}
and the corresponding minimum detection error probability is given by
\begin{equation}\label{eqq9}
\!\!{\xi ^\ast} = \left\{ {\begin{array}{*{20}{l}}
{q,\;{\kern 1pt} \;{\kern 1pt} \;{\kern 1pt} \;{\kern 1pt} \;{\kern 1pt} \;{\kern 1pt} \;{\kern 1pt} \;{\kern 1pt} \;{\kern 1pt} \;{\kern 1pt} \;{\kern 1pt} \;{\kern 1pt} \;{\kern 1pt} \;{\kern 1pt} \;{\kern 1pt} \;{\kern 1pt} \;{\kern 1pt} \;{\kern 1pt} \;{\kern 1pt} \;{\kern 1pt} \;{\kern 1pt} \;{\kern 1pt} \;{\kern 1pt} \;{\kern 1pt} \;\;{\kern 1pt}\;\;\;\;\;  q \le \frac{1}{{1 + \exp \left( {\frac{{{\lambda _{rw}}{{\left| {{h_{sw}}} \right|^2}}{P_s}}}{{{P_r}}}} \right)}},}\\
{(1 - q)\exp \left( { - \frac{{{\lambda _{rw}}{{ \left | {{h_{sw}}} \right|^2}}{P_s}}}{{{P_r}}}} \right),q > \frac{1}{{1 + \exp \left( {\frac{{{\lambda _{rw}}{{\left| {{h_{sw}}} \right|^2}}{P_s}}}{{{P_r}}}} \right)}}}.
\end{array}} \right.
\end{equation}
\end{theorem}

\begin{IEEEproof}
The detailed proof is presented in Appendix B.
\end{IEEEproof}


Based on Theorem~\ref{theorem2}, we note that the minimum detection error probability $\xi^\ast$ first increases and then decreases as $q$ increases. Meanwhile, we note that $\xi^\ast$ is affected by $p$ only through $q$. As such, it is obvious that maximizing $q$ by setting $p=1$ is not the best strategy to minimize $\xi^\ast$, which is different from the best strategy for conventional WPT to maximize the transmission throughput (e.g. \cite{Krikidis_single_Relay,accumulate,Bi_Ying}). In particular, we note that $p$ is an auxiliary parameter that can be used to prevent $q$ being exceedingly large regardless of the values of other parameters as per \eqref{add3}. Furthermore, we note that $\xi^\ast$ is affected by $P_s$ and $p$ determined at Sensor. From Sensor's point of view, the values of $P_s$ and $p$ should be carefully optimized in order to achieve the best covert communication performance with WPT, which is tackled in the following section.


\subsection{Covert Communications Design} \label{sec:3-3}
In order to jointly optimize $P_s$ and $p$ to maximize the covert communication performance, we first derive an explicit expression for the effective covert rate defined in \eqref{addd11} in the following proposition.

\begin{proposition}
The effective covert rate of the WP-CC system is given by
\begin{align}\label{eqq20}
{R_c} = qR\left( {1 - {P_\mathrm{out}}} \right) = qR\exp \left( { - \frac{{{\lambda _{sr}}\beta }}{{{P_s}}}} \right),
 \end{align}
where $\beta  = ({2^R} - 1)(\phi P_r+\sigma _b^2)$.
\end{proposition}

\begin{IEEEproof}
Based on the definition of the transmission outage probability, we have
\begin{align}\label{addd2}
{P_\mathrm{out}} &= \Pr \left\{ {{{\log }_2}\left( {1 + \frac{{{P_s}{{\left| {{h_{sr}}} \right|}^2}}}{{\phi P_r+\sigma _b^2}}} \right) < R} \right\} \nonumber \\
&= 1 - \exp \left( { - \frac{{{\lambda _{sr}}\beta }}{{{P_s}}}} \right),
\end{align}
which leads to the desired result in \eqref{eqq20}.
\end{IEEEproof}

As per \eqref{eqq20}, $R_c$ is affected by $P_s$ and $q$ simultaneously, where we recall that $P_s$ and $q$ have the non-linear relationship as given in \eqref{add3}. Meanwhile, we find that $R_c$ is affected by $p$ only through $q$. Considering the analysis in Section \ref{sec:3-1} and Section \ref{sec:3-2}, we need to further optimize $P_s$ and $p$ to achieve the best trade-off between the communication covertness and quality. Specifically, the optimization problem is given by
\begin{subequations}\label{add23}
\begin{align}
({\rm{\mathbf{P1}}}):&{\rm{   }}\mathop \mathrm{maximize}\limits_{P_s,\;p} \;\; \xi^\ast  \\
&{\rm{            }}\;{\rm{s.t.}}\quad {R_c} \ge R_m, \label{add6}\\
&{\rm{            }}\quad \quad\;\; {\rm{0}}\;{\rm{ < }}\;P_s \le P_m,\label{addd4}\\
&{\rm{            }}\quad \quad\;\; {\rm{0}}\;{\rm{ < }}\;p \le 1,\label{add8}
\end{align}
\end{subequations}
where $R_m$ is the minimum required value of the effective covert rate and $P_m$ is the maximum information transmit power. We note that \eqref{add6} represents a requirement on the quality-of-service (QoS). The feasibility condition of and solution to the optimization problem \eqref{add23} are tackled in the following theorem.

\begin{theorem}\label{theorem3}
The feasibility condition for the optimization problem \eqref{add23} is given by
 \begin{align}\label{add4}
{R_m} \le \frac{1}{{1 + {\mu _1}{\kappa _1}}}\exp \left( { - \frac{{{\lambda _{sr}}\beta }}{{{\kappa _1}}}} \right)R ,
\end{align}
under which the optimal values of $P_s$ and $p$ are given by
\begin{align}
P_s^ \ast  &= \mathop {\arg \;\max }\limits_{{s_1} < {P_s} \le \min \left( {{s_2},{P_m}} \right)} \;{f_2}\left( {{P_s}} \right),\label{add11}\\
{p^ \ast } &= \left\{ {\begin{array}{*{20}{l}}
{1,\;\;\;\;\;\;\;\;\;\;\;\;\;\;\;\;\;\;\;{\Theta _1} < 0,}\\
{\max \left( {{\kappa _2},{\kappa _3}} \right),\;{\Theta _1} \ge 0,}
\end{array}} \right.\label{addd1}
\end{align}
where
\begin{align}
\;&\mu_1=\frac{{{\lambda _{rs}}}}{{\eta {P_r}}},\;\; \mu_2=\frac{{{\lambda _{rw}}{{\left| {{h_{sw}}} \right|^2}}}}{{{P_r}}}, \nonumber \\
\;&\mu_3  = \frac{{{\lambda _{sr}}\beta }}{2}\left( {\sqrt {1{\rm{ + }}\frac{4}{{{\mu _{\rm{1}}}{\lambda _{sr}}\beta }}} {\rm{ + }}1} \right), \nonumber \\
\;&{\Theta _1} =\exp \left( {{\mu _2}{P_s^\ast}} \right) - {\mu _1}{P_s^\ast}, \nonumber \\
\;& {\Theta _2}=\frac{R}{{{R_m}}}\exp \left( { - \frac{{{\lambda _{sr}}\beta }}{{{P_s^\ast}}}} \right) - {\mu _1}{P_s^\ast} -1, \nonumber\\
\;& {\kappa _{\rm{1}}}{\rm{ = }}\min \left( {\mu_3 ,{P_m}} \right), \nonumber \\
\;&{\kappa _2} = \left\{ \begin{array}{l}
1,\;\;\;\;\;\;\;\;\;\;\;\;\;\;\;\;\;\;\;\;\;\;\;\;\;\;\;\;\;P_s^\ast =  - \frac{1}{{{\mu _2}}}{W_0}\left( { - \frac{{{\mu _2}}}{{{\mu _1}}}} \right),\\
\frac{1}{{2{\Theta _1}}}\left( {\sqrt {1{\rm{ + }}4{\Theta _1}}  - 1} \right),P_s^\ast  \ne   - \frac{1}{{{\mu _2}}}{W_0}\left( { - \frac{{{\mu _2}}}{{{\mu _1}}}} \right),
\end{array} \right.  \nonumber \\
\;& {\kappa _3} \!=\! \left\{ \begin{array}{l}
1,\;\;\;\;\;\;\;\;\;\;\;\;\;\;\;\;\;\;\;\;\;\;\;\;\;\;\;\;\;\;\;P_s^\ast = {\kappa _1},\\
\frac{1}{{2{\Theta _2}}}\left( {\sqrt {1{\rm{ + }}4{\Theta _2}}  - 1} \right),\;\;P_s^\ast \ne {\kappa _1},
\end{array} \right.\nonumber \\
& {f_1}({P_s}) = \frac{1}{{1{\rm{ + }}{\mu _1}{P_s}}}\exp \left( { - \frac{{{\lambda _{sr}}\beta }}{{{P_s}}}} \right)R - {R_m}, \nonumber  \\
\;&f_2(P_s)  = \left\{ {\begin{array}{*{20}{l}}
\!\!\!\!{\left( {1 \!\!-\!\! {f_3}({P_s})} \right)\exp \left( { - {\mu _2}{P_s}} \right),\;f_4(P_s) \le 0,}\\
\!\!\!{\frac{1}{{1 + {\mu _1}{P_s}}},\;\;\;\;\;\;\;\;\;\;\;\;\;\;\;\;\;\;\;\;\;\;\;\;\;\;\;
f_4(P_s) > 0,}
\end{array}} \right.  \nonumber \\
& {f_3}({P_s}) = \max \left( {\frac{{{R_m}}}{R}\exp \left( {\frac{{{\lambda _{sr}}\beta }}{{{P_s}}}} \right),\frac{1}{{1 + \exp \left( {{\mu _2}{P_s}} \right)}}} \right), \nonumber \\
\;&{f_4}\left( {{P_s}} \right) = {\mu _1}{P_s} - \exp \left( {{\mu _2}{P_s}} \right), \nonumber
\end{align}
$s_1$ and $s_2$ are the two solutions of $P_s$ to $f_1(P_s)\!=\!0$ with $s_1 \!\le\! s_2$, and ${W_k}\left(  \cdot  \right)$ is the $k$-th branch of the Lambert $W$ function.

\end{theorem}

\begin{IEEEproof}
The detailed proof is presented in Appendix C.
\end{IEEEproof}

Following Theorem~\ref{theorem3}, we note that the optimization problem in \eqref{add23} is infeasible, i.e., \eqref{add6} cannot be satisfied, when $P_r$ is extremely low or extremely high. We recall that $\beta  = ({2^R} - 1)(\phi P_r+\sigma _b^2)$ and, when $P_r\rightarrow 0$, we have $\mu_3 \rightarrow 0$ and thus $\kappa_1 = \mu_3$, which leads to the right hand side of \eqref{add4} approaching to 0 due to $\mu_1 \kappa_1 \rightarrow \infty$ and $\beta/\kappa_1 \rightarrow \infty$. Intuitively, the reason why \eqref{add6} cannot be satisfied when $P_r\rightarrow 0$ is that Sensor will not have enough energy for transmission when $P_r\rightarrow 0$. This is confirmed by that as $P_r\rightarrow 0$, the prior transmit probability $q$ derived in Theorem~\ref{theorem1} approaches to 0 as well.
Meanwhile, as $P_r\rightarrow \infty$, we have $\mu_3 \rightarrow \infty$ and thus $\kappa_1 = P_m$. Besides, we also have $\mu_1 \rightarrow 0$ and $\beta \rightarrow \infty$ as $P_r\rightarrow \infty$. This leads to $\frac{{\rm{1}}}{{{\rm{1 + }}{\mu _{\rm{1}}}{P_m}}} \rightarrow 1$ and  $\exp \left( { - \frac{{{\lambda _{sr}}\beta }}{{{P_m}}}} \right) \rightarrow 0$, which results in that the right hand side of \eqref{add4} approaches to 0. In addition, we note that most existing works on wireless communications with WPT always set $p=1$ to maximize the communication throughput (e.g., \cite{Krikidis_single_Relay,accumulate,Bi_Ying}), which means that Sensor transmits information immediately once the available energy is sufficient. However, as per Theorem~\ref{theorem3}, $p=1$ is not always the best strategy in the WP-CC system to maximize communication covertness subject to a constraint on communication quality.



To facilitate our comparison, we further present the feasibility condition and the solution to the optimization problem \eqref{add23} with $p=1$ in the following corollary.

\begin{corollary}\label{corollary1}
The feasibility condition for the case with $p=1$ is the same as \eqref{add4}, under which the optimal value of $P_s$ is given by
\begin{align}
P_s^ \ast= \min \left\{ {\max \left( {{s_1},{\kappa _4}} \right),\min \left( {{s_2},{P_m}} \right)} \right\}\label{add14},
\end{align}
where 
\begin{align}
{\kappa _4}{\rm{ = }}\left\{ \begin{array}{l}
\frac{1}{{2{\mu _{\rm{1}}}}}\left( {\sqrt {1 + \frac{{4{\mu _1}}}{{{\mu _2}}}}  - 1} \right),\;\frac{{{\mu _{\rm{1}}}}}{{{\mu _{\rm{2}}}}} \le \exp(1),\\
 - \frac{{\rm{1}}}{{{\mu _{\rm{2}}}}}{W_{\rm{0}}}\left( { - \frac{{{\mu _{\rm{2}}}}}{{{\mu _{\rm{1}}}}}} \right),\;\;\;\;\;\;\;\;\;\frac{{{\mu _{\rm{1}}}}}{{{\mu _{\rm{2}}}}}>\exp(1).
\end{array} \right. \nonumber
\end{align}
\end{corollary}

\begin{IEEEproof}
This proof follows along the lines of Appendix C and is omitted due to page limitation.
\end{IEEEproof}

Intuitively, as $R_m \rightarrow 0$ we should have $P_s \rightarrow 0$ in order to maximize the achievable
communication covertness \cite{fullduplex}. However, this is not the case for the WP-CC system with $p=1$. Following Corollary~\ref{corollary1}, we note that as $R_m \rightarrow 0$, the optimal transmit power for the WP-CC system with $p=1$ approaches to a non-zero constant value, which is independent of $R_m$. This is mainly due to the following fact.
We first note that as $P_s \rightarrow 0$ or $P_s \rightarrow \infty$, we have $\frac{1}{{1{\rm{ + }}{\mu _1}{P_s}}}\exp \left( { - \frac{{{\lambda _{sr}}\beta }}{{{P_s}}}} \right)R \rightarrow 0$ in the function ${f_1}({P_s})$ defined in Theorem~\ref{theorem3}. Recalling that $s_1>0$ and $s_2>0$ are the two solutions of $P_s$ to ${f_1}({P_s}) = 0$ with $s_1 \leq s_2$, we have $s_1 \rightarrow 0$ and $s_2 \rightarrow \infty$ as $R_m \rightarrow 0$. As $s_1 \rightarrow 0$ and $s_2 \rightarrow \infty$, we have $P_s^\ast = \min(\kappa_4,P_m)$ according to Corollary~\ref{corollary1}, where we note that neither $P_m$ nor $\kappa_4$ is a function of $R_m$. Intuitively, this is due to the fact that a smaller $P_s$ leads to a higher information transmit probability in the WP-CC system with $p=1$, which means that the communication activity becomes easier to be detected. This also demonstrates the necessity of using the proposed ATT protocol, where $p$ can be varied.


\section{Cable-Powered Covert Communications} \label{sec:4}

In this section, we tackle the optimal design of the CP-CC system. Specifically, we optimize the information transmit power $P_s$ and the prior transmit probability $q$ to maximize the communication covertness under a constraint on the effective covert rate. The obtained result serves as a performance benchmark for the WP-CC system, providing important system design insights.


As per \eqref{eqq9} and \eqref{eqq20}, we find that both of $\xi^\ast$ and $R_c$ are affected by $P_s$ and $q$ simultaneously. It is noted that we have $q=p$ in the CP-CC system, while the corresponding mapping in WP-CC system is non-linear as given in \eqref{add3}. Thus, we next optimize $P_s$ and $q$ to achieve a balance between the communication covertness and quality in the CP-CC system. Specifically, the optimization problem in the CP-CC system is given by
\begin{subequations}\label{add26}
\begin{align}
({\rm{\mathbf{P2}}}):&{\rm{   }}\mathop {\mathrm{maximize} }\limits_{P_s,\;q} \;\; \xi^\ast  \label{add15}\\
&{\rm{            }}\;{\rm{s.t.}}\quad {R_c} \ge R_m, \label{add16}\\
&{\rm{            }}\quad \quad\;\; {\rm{0}} \le P_s \le P_m, \label{add17}\\
&{\rm{            }}\quad \quad\;\; {\rm{0}} \le q \le 1, \label{add18}
\end{align}
\end{subequations}
where we recall that $\xi^\ast$ is given in \eqref{eqq9} and $R_c$ is given in \eqref{eqq20}. The feasibility condition of the optimization problem \eqref{add26} and its solution are presented in the following theorem.

\begin{theorem}\label{theorem4}
 The feasibility condition for the optimization problem \eqref{add26} is given by
\begin{align}\label{addd9}
{R_m} \le \exp \left( { - \frac{{{\lambda _{sr}}\beta }}{{{P_m}}}} \right)R,
\end{align}
under which the optimal values of $P_s$ and $q$ are given by
\begin{align}
P_s^\ast &=\mathop {\arg \;\max }\limits_{{\kappa _5} \le {P_s} \le {P_m}} \;{f_5}({P_s}), \label{add20}\\
{q^ \ast } &= \max \left( {{f_6}\left( {{P_s} = P_s^\ast} \right),{f_7}\left( {{P_s} = P_s^\ast} \right)} \right), \label{addd10}
\end{align}
where
\begin{align}
&  {\kappa _5} = \frac{{{\lambda _{sr}}\beta }}{{\ln \left( R \right) - \ln \left( {{R_m}} \right)}}, \nonumber \\
& {f_5}\left( {{P_s}} \right) = \left\{ \begin{array}{l}
\!\!\!\!\left( {1 \!\!-\!\! {f_6}\left( {{P_s}} \right)} \right)\exp \!\! \left( { - \frac{{{\lambda _{rw}}{{\left| {{h_{sw}}} \right|}^2}{P_s}}}{{{P_r}}}} \right),{f_6}\left( {{P_s}} \right) \!\!\ge\!\! {f_7}\left( {{P_s}} \right),\\
\!\!\!\!{f_7}\left( {{P_s}} \right),\;\;\;\;\;\;\;\;\;\;\;\;\;\;\;\;\;\;\;\;\;\;\;\;\;\;\;\;\;\;\;\;\;\;\;\;{f_6}\left( {{P_s}} \right) \!\!<\!\! {f_7}\left( {{P_s}} \right),
\end{array} \right. \nonumber \\
& {f_6}\left( {{P_s}} \right) = \frac{{{R_m}}}{R}\exp \left( {\frac{{{\lambda _{sr}}\beta }}{{{P_s}}}} \right),\nonumber \\
& {f_7}\left( {{P_s}} \right) = \frac{1}{{{\rm{1 + }}\exp \left( {\frac{{{\lambda _{rw}}{{\left| {{h_{sw}}} \right|}^2}{P_s}}}{{{P_r}}}} \right)}}. \nonumber
\end{align}
\end{theorem}

\begin{IEEEproof}
The proof follows along the lines of Appendix C  and is omitted for brevity.
\end{IEEEproof}

Following Theorem~\ref{theorem4}, we note that the optimization problem \eqref{add26} is infeasible when $P_r$ is exceedingly large. We recall that $\beta  = ({2^R} - 1)(\phi P_r+\sigma _b^2)$ and thus the right hand side of \eqref{addd9} approaches to 0 as $P_r \rightarrow \infty$, such that the feasibility condition \eqref{addd9} cannot be satisfied. Intuitively, this is due to that the CP-CC system suffers from high residual self-interference when $P_r$ is extremely large, which is confirmed by that the transmission outage probability derived in \eqref{addd2} approaches to 1 as $P_r \rightarrow \infty$. In addition, we note that $q=1/2$ is commonly adopted in the literature of covert communications (e.g. \cite{He_Biao,new_channel_uncertainty,Cognitive,Relay,zheng2019multi}), which is due to the assumption of unknown or equal prior probabilities for $H_0$
and $H_1$. However, as per Theorem~\ref{theorem4}, we know that $q=1/2$ is not always the best strategy in the CP-CC system. For example, if ${R_m} > R/2$, we have ${f_6}\left( {{P_s}} \right)>1/2$ and ${f_7}\left( {{P_s}} \right)<1/2$, which leads to $q^\ast >1/2$ as per \eqref{addd10}. This is due to the fact that adjusting $q$ in the CP-CC system can satisfy stricter requirements on the communication quality and thus can achieve higher communication covertness.

\section{Numerical Results}\label{sec:5}

In this section, we present numerical results to evaluate the performance of the considered WP-CC and CP-CC systems. Unless stated otherwise, we set the energy conversion efficiency as $\eta=0.8$, the duration of each time slot as $T=1$ s, the predetermined transmission rate as $R=1$ bit/s/Hz, the maximum information transmit power at Sensor as $P_m=5$ dBm, the AWGN variances at Receiver and Willie as $\sigma_b^2=\sigma_w^2=-60$ dBm, the self-interference cancellation coefficient as $\phi=-60$ dB, {and the mean values of the channel gains as $1/{\lambda _{sr}}=1/{\lambda _{rs}}=1/{\lambda _{rw}}={{\left| {{h_{sw}}} \right|^2}}={G_{ab}}{K^2}{\left( {\frac{d_{ab}}{{{d_0}}}} \right)^{ - \alpha }}$ \cite{Backscatter}, where $ab \in \left\{ {sr,rs,rw,sw} \right\}$, $K = \frac{c}{{4\pi f}}$ is a constant dependent upon the carrier frequency $f=470$ MHz, $c=3\times10^8$ m/s is the speed of light, $\alpha=2.7$ is the path loss exponent, $d_0=50$ m is the reference distance , $G_{ab}=12$ dB and $d_{ab}=100$ m are the combined transmitter-receiver antenna gain and the distance between node $a$ and node $b$, respectively.} In addition, we denote the optimal transmit probability as $p^\ast$ in the WP-CC system and $q^\ast$ in the CP-CC system, respectively. When the covert communications is infeasible, we set $\xi^\ast=0$, $P_s^\ast=0$ and $p^\ast=q^\ast=0$.

\begin{figure}
\centering
\label{Fig.3}
\includegraphics[width=0.5\textwidth]{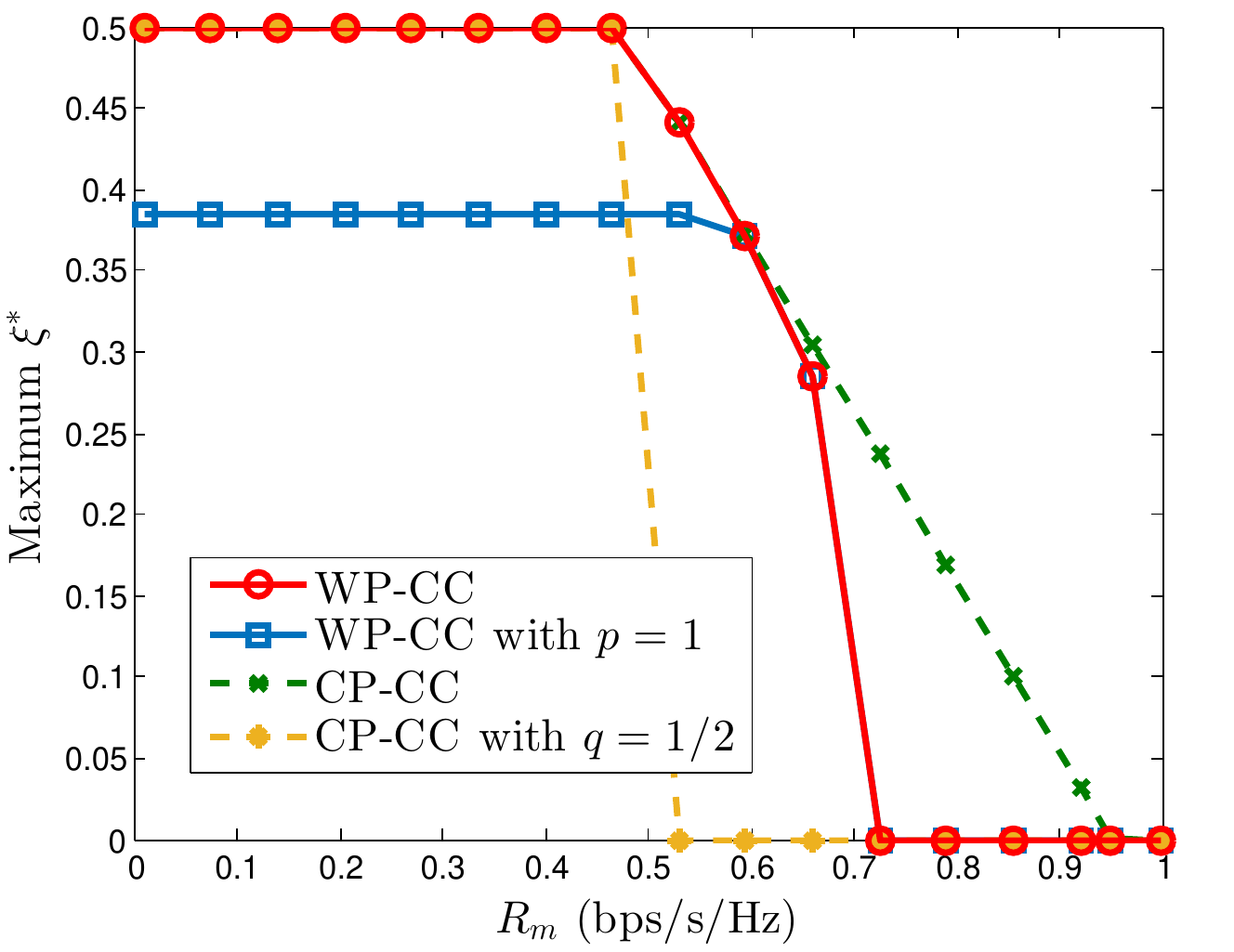}
\caption{The maximum $\xi^\ast$ versus the minimum effective covert rate $R_m$ with $P_r=30$ dBm.}
\end{figure}

In Fig. 3, we plot the maximum $\xi^\ast$ achieved by different systems versus $R_m$. In this figure, we first observe that the maximum $\xi^\ast$ in the WP-CC system with $p = 1$ is less than 0.4 when $R_m$ is relatively small, while other systems have the ability to achieve the perfect covertness (i.e., $\xi^\ast=\text{0.5}$). This demonstrates the necessity of varying the conditional prior  probability $p$ in the WP-CC system. Then, we observe that the maximum $\xi^\ast$ in the CP-CC system is an upper bound of the one achieved in the WP-CC system. In particular, we find that the WP-CC system can reach the upper bound for most values of $R_m$, while the maximum $\xi^\ast$ in the WP-CC system is lower than that of the upper bound for some other values of $R_m$, due to the stricter feasibility condition \eqref{add4}. Furthermore, we find that the WP-CC system even with $p=1$ can still acquire the positive covertness when the CP-CC system with $q=1/2$ cannot,
which is contributed by the intrinsic relationship between the information transmit probability and the information transmit power.

\begin{figure}
\centering
\label{Fig.4}
\includegraphics[width=0.5\textwidth]{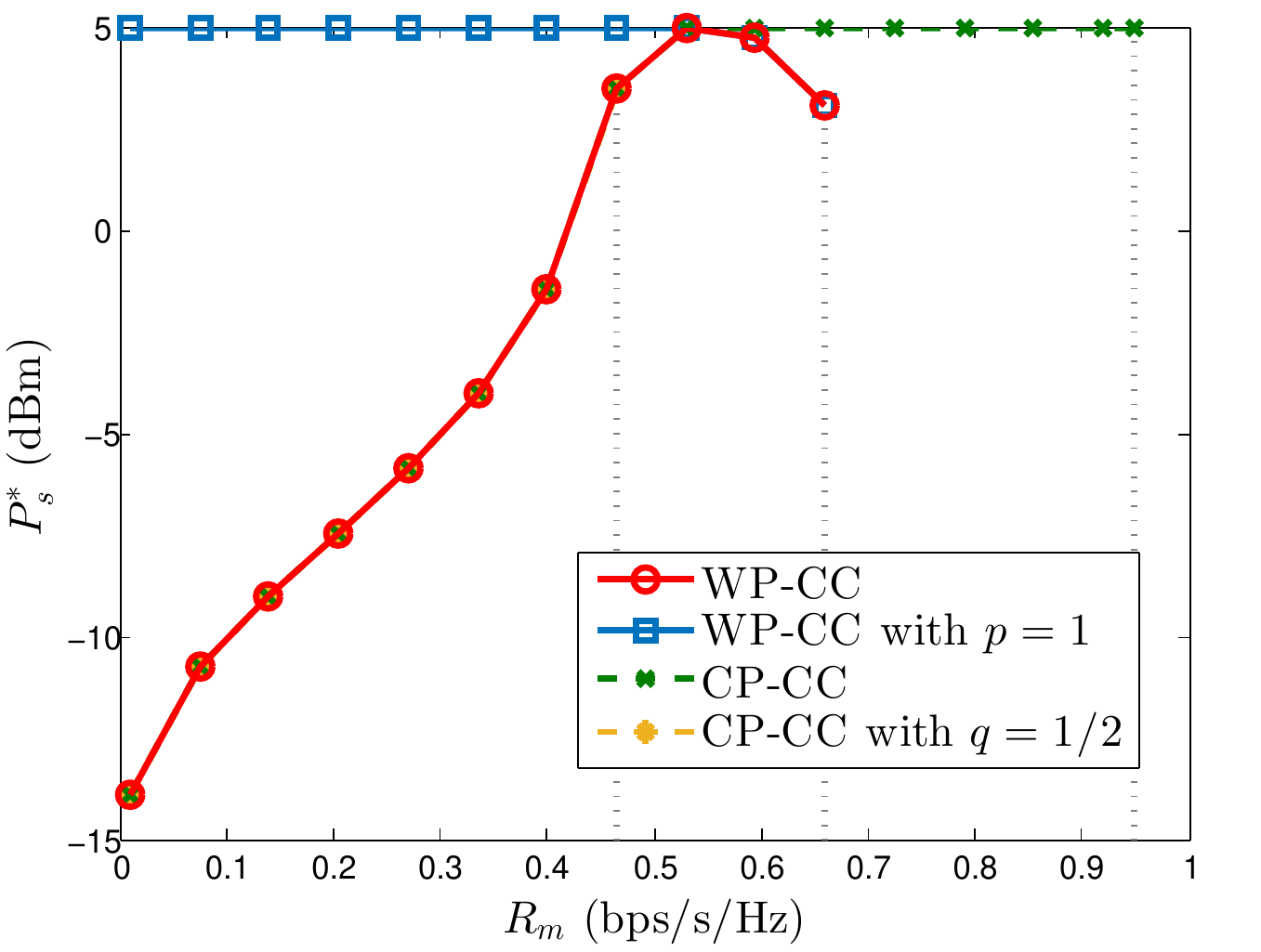}
\caption{The optimal information transmit power $P_s^\ast$ versus the minimum effective covert rate $R_m$ with $P_r=30$ dBm.}
\end{figure}

In Fig. 4, we plot the optimal information transmit power $P_s^\ast$ achieved by different systems versus $R_m$. {In this figure, we first observe that $P_s^\ast$ in the WP-CC system with $p=1$ remains unchanged at the maximum value and then decreases as $R_m$ increases.} This is due to the fact that a smaller $P_s$ leads to a higher transmit probability $q$ in the WP-CC system with $p=1$ (i.e., Sensor transmits signals more frequently), and thus the communication activity is easier to be detected. Meanwhile, increasing $P_s$ reduces $q$ and thus improves communication covertness. This is also confirmed by our analysis presented in Corollary~1 and the following discussions. In contrast, the WP-CC system with a varying $p$ can adopt the probabilistic ATT protocol (i.e., adjusting $p$) and the CP-CC systems can directly adopt a smaller information transmit probability to avoid such a negative effect. Furthermore, we find that the feasible regime of the WP-CC system is generally smaller than that of the CP-CC system, which is caused by the energy limitation in the WP-CC system.

\begin{figure}
\centering
\label{Fig.5}
\includegraphics[width=0.5\textwidth]{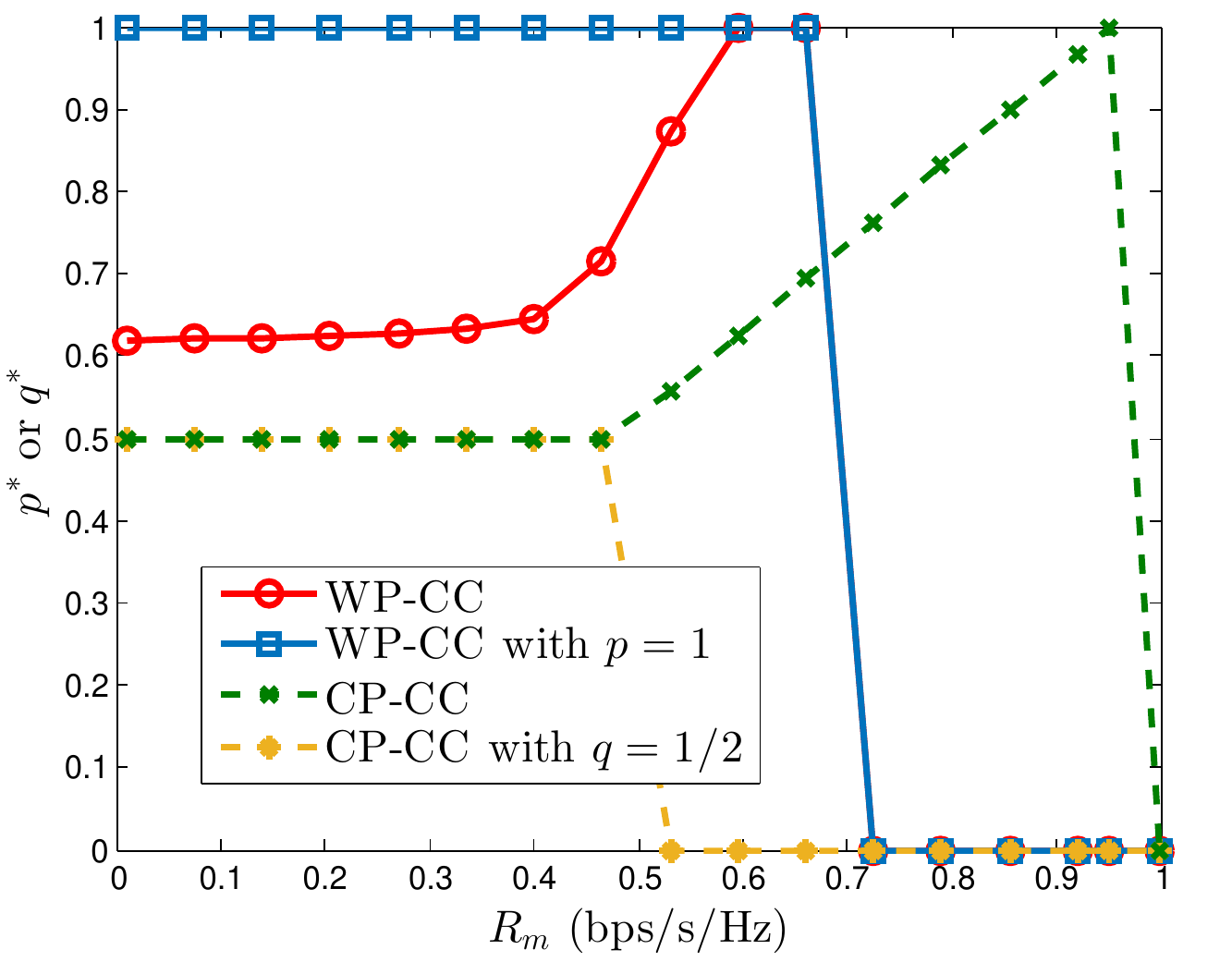}
\caption{The optimal transmit probability (i.e., $p^\ast$ in the WP-CC system or $q^\ast$ in the CP-CC system) versus the minimum effective covert rate $R_m$ with $P_r=30$ dBm.}
\end{figure}

In Fig. 5, we plot the optimal transmit probability (i.e., $p^\ast$ in the WP-CC system or $q^\ast$ in the CP-CC system) achieved by different systems versus $R_m$. In this figure, we first observe that the conventional ATT protocol (i.e., $p=1$) is not always the best strategy in the WP-CC system to achieve communication covertness. Meanwhile, we observe that the conventional transmit probability setting in covert communications (i.e., $q=1/2$) is also not always the best strategy in the CP-CC system to achieve communication covertness. Furthermore, we find that the feasible region of the CP-CC system is much larger than the corresponding special case with $q=1/2$. This is due to the fact that the CP-CC system can adopt a higher transmit probability to meet a stricter requirement on communication quality.

\begin{figure}
\centering
\label{Fig.6}
\includegraphics[width=0.5\textwidth]{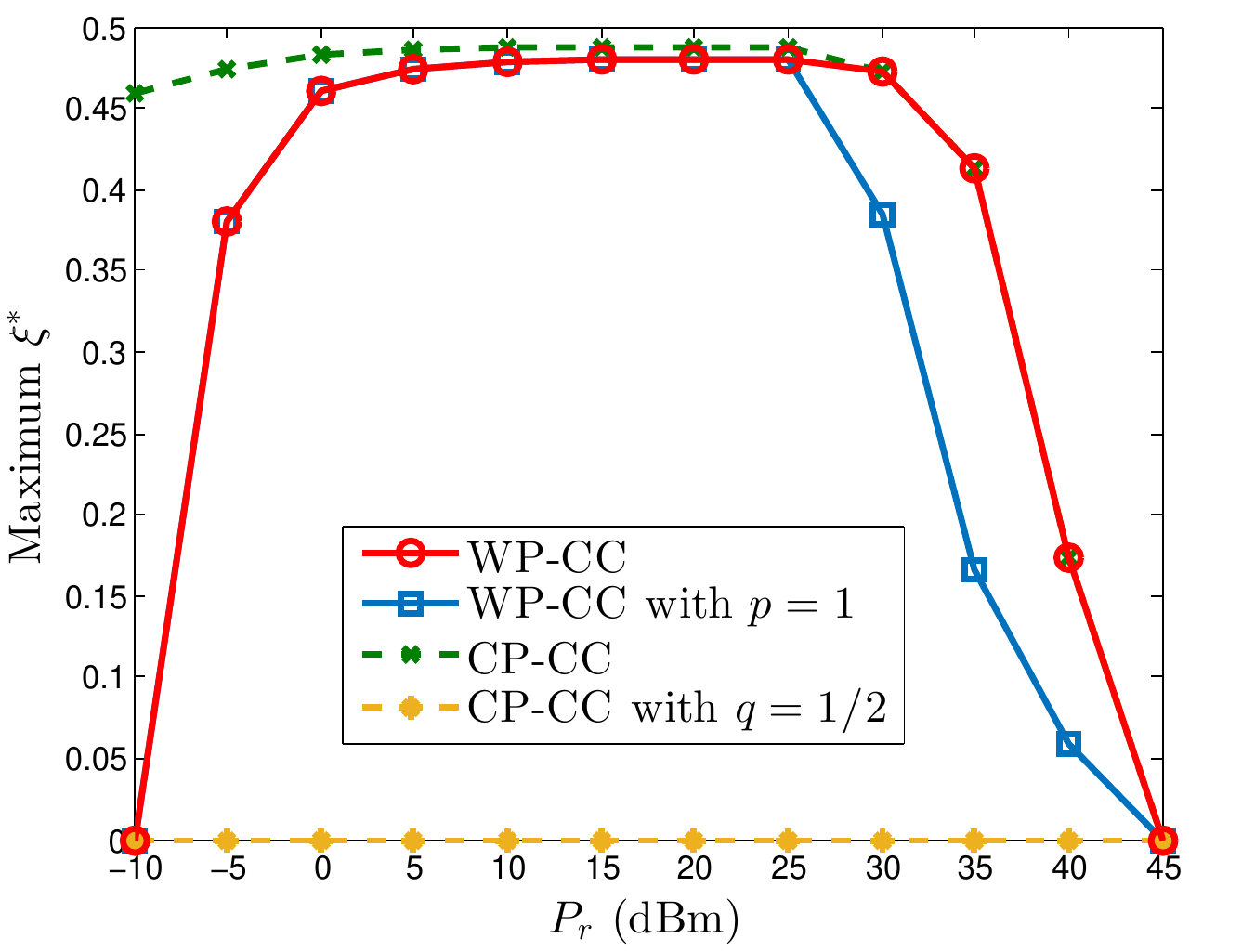}
\caption{The maximum $\xi^\ast$ versus the AN  transmit power $P_r$ with $R_m=0.5$ bit/s/Hz.}
\end{figure}

In Fig. 6, we plot the maximum $\xi^\ast$ achieved by different schemes versus the AN transmit power $P_r$ with $R_m=0.5$ bit/s/Hz. In this figure, we first observe that the maximum $\xi^\ast$ achieved by the WP-CC system first increases from 0 and then decreases to 0 as $P_r$ increases. The main reason is that it is hard to shield the communication activity or charge Sensor with sufficient amount of energy when the AN transmit power is low, while we would experience an exceedingly high self-interference at Receiver when the AN transmit power is high. In addition, we find that the performance of the WP-CC system is exactly the same as that of the CP-CC system when the AN transmit power is in the high regime as the energy insufficiency no longer limits the performance.
However, we also observe that the WP-CC system with $p=1$ cannot achieve the same performance as the CP-CC system. As such, the aforementioned two observations demonstrate the benefits of using the proposed ATT protocol in the WP-CC system.

\section{Conclusion} \label{sec:6}
In this paper, we developed a framework to optimally design the WP-CC system with a probabilistic ATT protocol, in which the transmitter deliberately continued to harvest energy rather than sending signals with a certain probability even when the available energy was sufficient for information transmission. Specifically, we first determined the non-linear relationship between the information transmit power and the prior transmit probability at the energy-constrained transmitter, which jointly affected the achievable communication covertness quantified by the minimum detection error probability. Under a constraint on communication quality, we further provided design guidelines to maximize the communication covertness for the WP-CC system and the CP-CC system as a benchmark. Our examination showed that the proposed probabilistic ATT protocol can achieve higher communication covertness than the traditional ATT protocol (i.e., the transmitter transmits information once the available energy is sufficient).
\begin{appendices}
\section{Proof of Lemma 1}
It is noted that the available energy at Sensor at the start of the $k$-th time slot is equal to the residual energy at the end of the $(k-1)$-th time slot. In order to derive the pdf of ${E_0}\left( t_k \right)$, we need to analyze the pdf of the corresponding residual energy, which can be divided into the following three cases.

\noindent
\textit{Case \uppercase\expandafter{\romannumeral1}}: ${t_k} \le {t'_1}$

For this case, $E_a^{\left( 0 \right)}=0$ holds. Hence, Sensor only decides to transmit information until $E_a^{\left( k \right)} \ge P_sT $. Since ${\left| {{h_{rs}}} \right|^2}$ is exponentially distributed with mean $1/\lambda_{rs}$, $E_h^{\left( k \right)}=\eta {P_r}{\left| {{h_{rs}}} \right|^2}T$ is exponentially distributed with mean $1/{\lambda _{{E_h}}}$ and the pdf is given by
\begin{equation}
f_{E_h^{\left( k \right)}} \left( z \right){\rm{ = }}\lambda _{{E}}\exp \left( { - \lambda _{{E}}z} \right){\rm{,}}\;z{\rm{ > }}0,
\end{equation}
where ${\lambda _{{E}}}=\frac{{{\lambda _{rs}}}}{{\eta {P_r}T}}$. Based on the memoryless property of the exponential distribution \cite[Definition 2.3]{Discrete_Book}, $E_a^{\left( k \right)}$ still follows the exponential distribution with the mean $1/{\lambda _{{E}}}$.

\noindent
\textit{Case \uppercase\expandafter{\romannumeral2}}: ${t'_1}{\rm{ < }}{t_k} \le {t'_2}$

If ${t'_2} - {t'_1} > 1$, Sensor still needs to harvest energy until $E_a^{\left( k \right)} \ge P_sT $. Thus, the distribution of $E_a^{\left( k \right)}$ is the same as the one in \textit{Case \uppercase\expandafter{\romannumeral1}}, which follows the exponential distribution with mean $1/{\lambda _{{E}}}$.

If ${t'_2} - {t'_1} = 1$, it means $E_a^{\left( M_1 \right)} \ge P_sT $. Then, the conditional cumulative distribution function (cdf) of $E_a^ {(M_2) } $ is given by
\begin{align}\label{eqq1}
{F_{E_a^{\left( {{M_2}} \right)}}}\left( z \right) &= \Pr \left( {E_a^{\left( {{M_2}} \right)} - {P_s}T < z|E_a^{\left( {{M_1}} \right)} \ge {P_s}T} \right)\nonumber \\ &= 1 - \exp \left( { - {\lambda _E}z} \right),z > 0,
\end{align}
and the corresponding pdf of $E_a^ {(M_2) } $ is given by
\begin{align}
{f_{E_a^{\left( {{M_2}} \right)}}}\left( z \right){\rm{ = }}{\lambda _E}\exp \left( { - {\lambda _E}z} \right),\;{\kern 1pt} z > 0.
\end{align}
It means that the distribution of $E_a^ {(M_2) } $ still follows the exponential distribution with mean $1/{\lambda _{{E}}}$. Combining above results, we conclude that $E_a^ {(k) } $ still follows the exponential distribution with mean $1/{\lambda _{{E}}}$ when ${t'_1}<{t_k} \le {t'_2}$.

\noindent
\textit{Case \uppercase\expandafter{\romannumeral3}}:
${t_k}{\rm{ > }}{t'_2}$

Based on the analysis for \textit{Case \uppercase\expandafter{\romannumeral1}} and \textit{Case \uppercase\expandafter{\romannumeral2}}, we can acquire the same result for $E_a^{\left( k \right)}$ when ${t_k}>{t'_2}$. In summary, $E_a^{\left( k \right)}$ in any ${t_k}$ follows the exponential distribution with mean $1/{\lambda _{{E}}}$.

\section{Proof of Theorem 2}
In order to determine the optimal detection threshold that minimizes detection error probability, we need
to tackle the following optimization problem
\begin{equation}
\mathop {\text{minimize} }\limits_\tau \;\; \xi,
\end{equation}
where the expression of $\xi$ is given in \eqref{eqq10}. Then, we tackle the solution to this optimization problem in the following three cases depending on the values of $\tau$ given in \eqref{eqq10}.

\noindent
\textit{Case \uppercase\expandafter{\romannumeral1}}: $\tau  \le \sigma _w^2$

Here, $\xi=1-q$ cannot be minimized by any $\tau$.

\noindent
\textit{Case \uppercase\expandafter{\romannumeral2}}: $\sigma _w^2 < \tau  \le \sigma _w^2 + {P_s}{{\left| {{h_{sw}}} \right|^2}}$

The first derivative of $\xi$ w.r.t. $\tau$ is given by
\begin{equation}
\frac{{\partial \xi }}{{\partial \tau }}{\rm{ = }} - {\lambda _{rw}}\left( {1 - q} \right)\exp \left( { - \frac{{{\lambda _{rw}}\left( {\tau  - \sigma _w^2} \right)}}{{{P_r}}}} \right) \le {\rm{0}}.
\end{equation}
Thus, $\xi$ is a monotonically decreasing function of $\tau$ in this case. The corresponding optimal value of $\tau$ is ${\tau ^\ast}=\sigma _w^2 + {P_s}{{\left| {{h_{sw}}} \right|^2}}$.

\noindent
\textit{Case \uppercase\expandafter{\romannumeral3}}: $\tau  > \sigma _w^2 + {P_s}{{\left| {{h_{sw}}} \right|^2}}$

The first derivative of $\xi$ w.r.t. $\tau$ is given by
\begin{align}\label{addd3}
\frac{{\partial \xi }}{{\partial \tau }} &= \left( { - {\lambda _{rw}}} \right)\left[ {\left( {1 - q} \right)\exp \left( { - \frac{{{\lambda _{rw}}\left( {\tau  - \sigma _w^2} \right)}}{{{P_r}}}} \right)} \right. \nonumber \\
& - \left. {q\exp \left( { - \frac{{{\lambda _{rw}}\left( {\tau  - \sigma _w^2 - {P_s}{{\left| {{h_{sw}}} \right|}^2}} \right)}}{{{P_r}}}} \right)} \right].
\end{align}
When $q > \frac{1}{{1 + \exp \left( {\frac{{{\lambda _{rw}}{{\left| {{h_{sw}}} \right|^2}}{P_s}}}{{{P_r}}}} \right)}}$, $\xi$ is a monotonically increasing function of $\tau$ as for \eqref{addd3} and the corresponding optimal value of $\tau$ is $\tau^\ast=\sigma _w^2 + {P_s}{{\left| {{h_{sw}}} \right|^2}}$. Otherwise, $\xi$ is a monotonically decreasing function of $\tau$ in this case and the corresponding optimal value of $\tau$ is $\tau^\ast= + \infty $.

Consequently, we obtain the overall optimal $\tau^\ast$ given in \eqref{eqq11} and the corresponding minimum detection error probability $\xi^\ast$ given in \eqref{eqq9}.

\section{Proof of Theorem 3}

We first determine the feasibility condition of the optimization problem \eqref{add23}. It is noted that $P_s$ and $p$ are independent to each other in \eqref{eqq20}, and thus we can analyze the monotonicity of $R_c$ w.r.t. $P_s$ and $p$, separately. Specifically, the monotonicity of $R_c$ w.r.t. $p$ depends on $\frac{{{\rm{1}} - p}}{{{p^2}}}$, which is a monotonically decreasing function of $p$ as per \eqref{add98}. Considering the constraint \eqref{add8}, we find that $R_c$ is a monotonically increasing function of $p$ and the maximum value of $R_c$ for a given $P_s$ can be acquired when $p=1$. Then, we further analyze the monotonicity of $R_c(p\!=\!1)$ w.r.t. $P_s$. The corresponding first derivative is given by
\begin{align}
\!\!\!\!\!\frac{{\partial {R_c(p=1)}}}{{\partial {P_s}}} &=\nonumber \\
&\!\!\!\!\!\!\!\!\!\!\!\!\!\!\!\!\!\!\!\! \frac{R}{{1 + {\mu _1}{P_s}}} \left( {\frac{{{\lambda_{sr} \beta}}}{{P_s^2}} - \frac{{{\mu _1}}}{{1 + {\mu _1}{P_s}}}}  \right)  \exp \left( { - \frac{{{\lambda_{sr} \beta}}}{{{P_s}}}} \right). \label{add9}
\end{align}
After some algebra manipulations, we find that the monotonicity of $R_c(p\!=\!1)$ w.r.t. $P_s$ depends on $ - P_s^2 + {\lambda _{sr}}\beta {P_s} + \frac{{{\lambda _{sr}}\beta }}{{{\mu _1}}}$. Based on the properties of univariate quadratic equation, we can conclude that if $0 < {P_s} \le {\frac{{{\lambda _{sr}}\beta }}{2}\left( {\sqrt {1{\rm{ + }}\frac{4}{{{\mu _{\rm{1}}}{\lambda _{sr}}\beta }}} {\rm{ + }}1} \right)}$, $R_c(p\!=\!1)$ is a monotonically increasing function of $P_s$. Otherwise, $R_c(p\!=\!1)$ is a monotonically decreasing function of $P_s$. Considering the constraint \eqref{addd4}, the maximum value of $R_c$ is $R_c(p=1, P_s=\kappa_1)$. In order to further satisfy the constraint \eqref{add6}, $R_c(p\!=\!1,P_s\!=\!\kappa_1)$ must be no less than $R_m$, which leads to the feasibility condition given in \eqref{add4}.

On the other hand, if the feasibility condition holds, we can further optimize $P_s$ and $p$ based on the monotonicity of $\xi^\ast$ w.r.t. $P_s$ and $p$. To ease further analysis, we denote $s_1$ and $s_2$ as the two solutions of $P_s$ to $f_1(P_s)=0$ with $s_1 \!\le\! s_2$. We note that $P_s$ and $p$ are independent in the expression for $\xi^\ast$ given in \eqref{eqq9}, and thus we can also analyze the monotonicity of $\xi^\ast$ w.r.t. $P_s$ and $p$, separately. Specifically, as per \eqref{add3}, we find that $q$ is a monotonically increasing function of $p$. Due to the constraint \eqref{add8}, for a given $P_s$, the maximum value of $q$ is $q=\frac{1}{{1 + {\mu _1}{P_s}}}$ achieved by $p\!=\!1$. Based on the two cases of $\xi^\ast$ given in \eqref{eqq9}, for a given $P_s$, we optimize $p$ in the following two cases.

\noindent
\textit{Case \uppercase\expandafter{\romannumeral1}}: ${f_4}\left( {{P_s}} \right) \le 0$

In this case, $\xi^\ast$ is a monotonically increasing function of $q$ if $q < \frac{1}{{1 + \exp \left( {{\mu _2}{P_s}} \right)}}$. Otherwise, $\xi^\ast$ is a monotonically decreasing function of $q$. Consequently, $\xi^\ast$ first increases then decreases with $q$. As per \eqref{add3}, $q$ is a monotonically increasing function of $p$. As such, the maximum value of $\xi^\ast$ can be achieved when $p=\kappa_2'$, where ${\kappa _2'} = {\kappa _2}\left( {P_s^* = {P_s}} \right)$ is the solution of $p$ to $q(p) = \frac{1}{{1 + \exp \left( {{\mu _2}{P_s}} \right)}}$ with ${\rm{0}}\;{\rm{ < }}\;p \le 1$. In order to meet the constraint \eqref{add6}, $p$ must be no less than $\kappa_3'$, where $\kappa_3'=\kappa_3(P_s^\ast=P_s)$ is the solution of $p$ to $R_c(p)=R_m$ with ${\rm{0}}\;{\rm{ < }}\;p \le 1$. Thus, for a given $P_s$, the optimal value of $p$ and the optimal value of $\xi^\ast$ are $p=\max(\kappa_2',\kappa_3')$ and $\xi^\ast=\left( {1 - {f_3}({P_s})} \right)\exp \left( { - {\mu _2}{P_s}} \right)$, respectively.

\noindent
\textit{Case \uppercase\expandafter{\romannumeral2}}: ${f_4}\left( {{P_s}} \right) > 0$

In this case, $\xi^\ast$ is a monotonically increasing function of $q$. As per \eqref{add3}, $q$ is a monotonically increasing function of $p$. Consequently, $\xi^\ast$ is also a monotonically increasing function of $p$. Due to the constraint \eqref{add8}, the maximum value of $\xi^\ast$ can be achieved when $p=1$. We note that the constraint \eqref{add6} must be satisfied if $f_4(P_s) >0 $. Thus, for a given $P_s$, the optimal value of $p$ and the corresponding optimal value of $\xi^\ast$ are given by $p=1$ and $\xi^\ast=\frac{1}{{1 + {\mu _1}{P_s}}}$, respectively.

Combining the analysis of \textit{Case \uppercase\expandafter{\romannumeral1}} and \textit{Case \uppercase\expandafter{\romannumeral2}}, for a given $P_s$, the optimal value of $\xi^\ast$ is given by $f_2(P_s)$. For a given $p$, considering the constraints \eqref{add6} and \eqref{addd4}, the feasible region of $P_s$ is ${P_s} \in \left[ {{s_1},\min \left( {{s_2},{P_m}} \right)} \right]$. Then, the optimization problem \eqref{add23} degrades to
\begin{subequations}\label{add25}
\begin{align}
({\rm{\mathbf{P1.1}}}):&{\rm{   }}\mathop {\mathrm{maximize} }\limits_{P_s} \;\; f_2(P_s)  \\
&{\rm{            }}\;{\rm{s.t.}}\quad s_1 \le P_s \le \min(s_2,P_m).
\end{align}
\end{subequations}
As such, the optimal value of $P_s$ can be acquired by an simple one-dimensional numerical search, which is denoted as $P_s^\ast$ given in \eqref{add11}. By substituting $P_s^\ast$ into the conditions of \textit{Case \uppercase\expandafter{\romannumeral1}} and \textit{Case \uppercase\expandafter{\romannumeral2}}, we can acquire the optimal value of $p$, which is denoted as $p^\ast$ given in \eqref{addd1}.

\end{appendices}

\small{\bibliographystyle{IEEEtran}}
\bibliography{double_column}
\end{document}